\documentclass{article}

\usepackage[preprint]{neurips_2019}

\usepackage[utf8]{inputenc} 
\usepackage[T1]{fontenc}    
\usepackage{hyperref}       
\usepackage{url}            
\usepackage{booktabs}       
\usepackage{amsfonts,amsmath}       
\usepackage{nicefrac}       
\usepackage{microtype}      
\usepackage{graphicx}
\usepackage{subfigure}

\usepackage{color}

\title{Site-specific graph neural network for predicting protonation energy of oxygenate molecules}

\author{%
  Romit Maulik \\
  Argonne Leadership Computing Facility\\
  Argonne National Laboratory\\
  Lemont, IL 60439 \\
  \texttt{rmaulik@anl.gov} \\
   \And
   Rajeev S. Assary  \\
   Material Science Division \\
   Argonne National Laboratory\\
   Lemont, IL 60439 \\
   \texttt{assary@anl.gov} \\
   \And
   Prasanna Balaprakash  \\
   Mathematics and Computer Science Division \\
   Argonne National Laboratory\\
   Lemont, IL 60439 \\
   \texttt{pbalapra@anl.gov} \\
}

\begin{document}

\maketitle

\begin{abstract}

Bio-oil molecule assessment is essential for the sustainable development of chemicals and transportation fuels. These oxygenated molecules have adequate carbon, hydrogen, and oxygen atoms that can be used for  developing new value-added molecules (chemicals or transportation fuels). One motivation for our study stems from the fact that a liquid phase upgrading using mineral acid is a cost-effective chemical transformation. In this chemical upgrading process, adding a proton (positively charged atomic hydrogen) to an oxygen atom is a central step. The protonation energies of oxygen atoms in a molecule determine the thermodynamic feasibility of the reaction and likely chemical reaction pathway. A quantum chemical model based on coupled cluster theory is used to compute accurate thermochemical properties such as the protonation energies of oxygen atoms and the feasibility of protonation-based chemical transformations. However, this method is too computationally expensive to explore a large space of chemical transformations. We develop a graph neural network approach for predicting protonation energies of oxygen atoms of hundreds of bioxygenate molecules to predict the feasibility of aqueous acidic reactions. Our approach relies on an iterative local nonlinear embedding that gradually leads to global influence of distant atoms and a output layer that predicts the protonation energy. Our approach is geared to site-specific predictions for individual oxygen atoms of a molecule in comparison with commonly used graph convolutional networks that focus on a singular molecular property prediction. We demonstrate that our approach is effective in learning the location and magnitudes of protonation energies of oxygenated molecules.
\end{abstract}

\section{Introduction}

Bio-oils, obtained from biomass such as wood and biowaste, play a critical role in producing new sustainable chemicals and fuels. Oxygenated molecules are derivatives of bio-oils with carbon, hydrogen, and oxygen atoms. Development of new value-added molecules (chemicals or transportation fuels) from naturally abundant oxygenates requires complex chemical transformations. 
Reactions using water in a mineral acid medium are cost and energy efficient for chemical transformations. The simplest step of acid medium reaction in water is protonation, where a proton from the mineral acid is absorbed on an electronegative part of the molecule, often the oxygen site. The protonation reaction of a hydroxyl group (OH) often leads to a dehydration reaction, which is a dominant reaction step in the biomass conversion reactions. Moreover, the thermodynamic and kinetic feasibility of the dehydration reaction can be predicted with the understanding of the protonation energy. Accurate estimates of the feasibility of proton-catalyzed reaction in aqueous media of possible intermediates are extremely costly and difficult, however. The state-of-the-art approach for protonation energy calculation is based on quantum chemical models. For example, G4MP2 [1,2], a model based on coupled cluster theory, is predicted to be accurate up to $\approx$1 kcal/mol for atomization energy compared with gas phase experiments. A disadvantage of G4MP2 is that it is computationally expensive. For example,  a molecule with 15 heavy atoms  may require 48 hours on a moderately sized cluster with hundreds of modern CPU nodes. For molecules with more than 15 heavy atoms, performing accurate quantum chemical computations is impossible. To that end, we develop a graph neural network approach that learns from the high-fidelity G4MP2 bio-oil database and predicts protonation energies of oxygenates.

\section{Graph neural network for site-specific protonation energy prediction}

Here, we describe our graph convolutional neural network (GCN) for site-specific protonation energy predictions. Our supervised learning approach requires training examples of arbitrarily sized molecular graphs with input and output features defined on each atom. Consequently, this learning necessitates predictions at each atom and correspondingly requires the specification of a modified regression approach for site-specific accuracy. Our approach is different from many deployments of GCNs, which are generally tasked with the prediction of a global property for the molecular graph. For an early example, in [3] a GCN embedding strategy was used to encode molecules into fixed-length fingerprint vectors. These vectors could then be mapped to a molecular properties. The study in [4] was similar to [3] where a GCN was used to map from a crystalline graph to its material properties. In [5] a molecular graph approach was proposed in which atomic, bond, and global state information was used to obtain a map to respective targets. The global state could represent some property such as reaction temperature affecting a transformation from inputs to targets. In the absence of global state information, however, and lacking the ability to deal with arbitrary molecular sizes, this approach is unsuitable for our task. In [6], GCNs were used to find reaction centers between pairs of arbitrary-sized molecules using a process of attention. Since this addressed our need for handling arbitrary shaped inputs, a similar embedding procedure was utilized before converting this problem to our regression task.

The inputs and response values for the regression task are outlined in Table \ref{Table1}. In addition to atomic properties, we encode bond features and connectivity information for appropriate information of an atomic neighborhood. Our targets are given by protonation energies at oxygen atoms and zero magnitudes at other locations. Our raw data containing structural information is available in xyz format (i.e., atomic numbers and their positions in 3D space). Each xyz representation for a molecule is pipelined through RDKit, an open-source cheminformatics software, for encoding molecular descriptors.

\begin{table}[]
\centering
\begin{tabular}{|c|c|c|}
\hline
\textbf{Atomic Feature}     & \textbf{Possibilities}    & \textbf{Representation} \\ \hline
Element              & H, C, O, Unknown  & One-hot        \\ \hline
Degree               & 0,1,2,3,4,5       & One-hot        \\ \hline
Explicit valence    & 1,2,3,4,5,6       & One-hot        \\ \hline
Implicit valence     & 0,1,2,3,4,5       & One-hot        \\ \hline
Aromatic             & 0,1               & Bool           \\ \hline
Number of neighbors  & -                 & Integer        \\ \hline
List of neighbors    & All atoms         & One-hot        \\ \hline
\textbf{Bond Feature}     & \textbf{Possibilities}    & \textbf{Representation} \\ \hline
Bond type    & Single, double, triple & One-hot       \\ \hline
Aromatic     & 0,1                    & Bool           \\ \hline
Conjugated   & 0,1                    & Bool          \\ \hline
In ring      & 0,1                    & Bool          \\ \hline
Connecting atoms    & All atoms       & One-hot        \\ \hline
\textbf{Response value}     & \textbf{Possibilities}    & \textbf{Representation} \\ \hline
Protonation energy  & -               & Real valued \\ \hline
\end{tabular} \vspace{0.2cm}
\caption{List of atom-specific input and output of our graph neural network.}
\label{Table1}
\end{table}

Our GCN approach is based on [6,7]. We note that the input features are all molecule specific and that their dimensions scale with the total number of atoms. To that end, we utilize a linear embedding into a uniform space through a matrix multiplication. This matrix is arbitrarily sized for each molecule and is initialized with a random-normal initialization. This arbitrary nature of embedding also implies that the elements of this matrix remain ``nontrainable'' during the learning process. The result of the initial encoding leads to a data sample that has the same dimensions for the features at each atom across our collected dataset. We  denote this initial encoding as 
\begin{align}
\mathbf{h}_v^0 = W^{ip} \mathbf{a}_v,
\end{align}
where $\mathbf{a}_v \in \mathbb{R}^{N_f}$ are $N_f$ input features (both atomic and bond) for an atom $v$ and $W^{ip} \in \mathbb{R}^{N_e \times N_f}$ is the nontrainable weight matrix that maps to the initial encoding $\mathbf{h}_v^0 \in \mathbb{R}^{N_f}$. Following this linear encoding of the input features at each atom, an iterative nonlinear embedding is performed for incorporating neighboring influences as follows:
\begin{align}
\mathbf{h}_{v}^{l}=\sigma\left(\mathbf{U}_{1} \mathbf{h}_{v}^{l-1}+\mathbf{U}_{2} \sum_{u \in N(v)} \sigma\left(\mathbf{V} \mathbf{h}_{u}^{l-1}\right)\right),
\end{align}
where $l \in \mathbb{R}^1$ indicates a particular layer of the network; $\sigma: \mathbb{R}^{N_e} \rightarrow \mathbb{R}^{N_e}$ is a nonlinear activation function; $\mathbf{U}_1, \mathbf{U}_2, \mathbf{V} \in \mathbb{R}^{N_e \times N_e}$ are trainable linear operations (i.e., weights and biases) shared between layers; and $N(v)$ stands for the neighborhood of atoms near $v$. This iterative embedding leads to the gradual influence of global information at a particular atom $v$ with increasing $l$. Following a comprehensive globally nonlinear encoding, a final output encoding is determined for ensuring comparison with protonation data at each atom (after a sufficient depth $l=L$) given by
\begin{align}
\mathbf{h}_{v}^{L}=\mathbf{U}_{1}^{L} \mathbf{h}_{v}^{L-1}+\mathbf{U}_{2}^{L} \sum_{u \in N(v)} \sigma\left(\mathbf{V}^{L} \mathbf{h}_{u}^{L-1}\right),
\end{align}
where $\mathbf{U}_{1}^{L}, \mathbf{U}_{2}^{L}, \mathbf{V}^{L} \in \mathbb{R}^{1 \times N_e}$ represent learnable parameters for leading to a one-output prediction at each atom given by $\mathbf{h}_v^L \in \mathbb{R}^1$.

The output features (i.e., the predictions) given by $\mathbf{h}_v^L$ are assessed against targets given by $\mathbf{t}_u^L \in \mathbb{R}^{N_1}$. For optimizing the trainable operations (i.e., $\mathbf{U}_1, \mathbf{U}_2, \mathbf{V}, \mathbf{U}^{L}_1, \mathbf{U}^{L}_2$ and $\mathbf{V}^{L}$), a loss function for each data sample $j$ is specified by
\begin{align}
F_j = \sum_{i=0}^{N_u} \lambda_i \left( \mathbf{h}_i^L - \mathbf{t}_i^L \right)^2 ,
\end{align}
where $N_u \in \mathbb{Z}^1$ is the number of atoms in the molecule and $\lambda_i$ is a sparsity promoting multiplier given by
\begin{align}
\lambda_i = 
\begin{cases}
    1,& \text{if } t_i = 0\\
    \lambda,  & t_i > 0
\end{cases}
\end{align}
with a total loss value for a batch given by
\begin{align}
F = \frac{1}{N_b} \sum_{j=0}^{N_b} F_j ,
\end{align}
where $N_b \in \mathbb{Z}^1$ is the number of samples in a batch. 

\section{Experimental results}


The protonation energy of molecular sites was generated from $\approx 2,000$ G4MP2 calculations using the Gaussian 16 software. This dataset was split into 70\% for training the GCN and 30\% for testing. We trained our GCN with the Adam optimizer using a learning rate of 1e\textsuperscript{-3} and a learning rate decay multiplier of 0.9 every 50 batch gradient updates. We ran the training for 2000 epochs and observed a sparsity-based root mean square error (RMSE) of 1030.8 across training. We observed that training and validation losses did not change beyond this range of training for further gradient updates. Our network hyperparameters were $N_e = 40, L = 5, N_b = 40, \lambda = 100, and  \sigma: \text{ReLU}$. 

\begin{figure}[ht]
\centering
\mbox{
\subfigure{\includegraphics[width=0.33\textwidth]{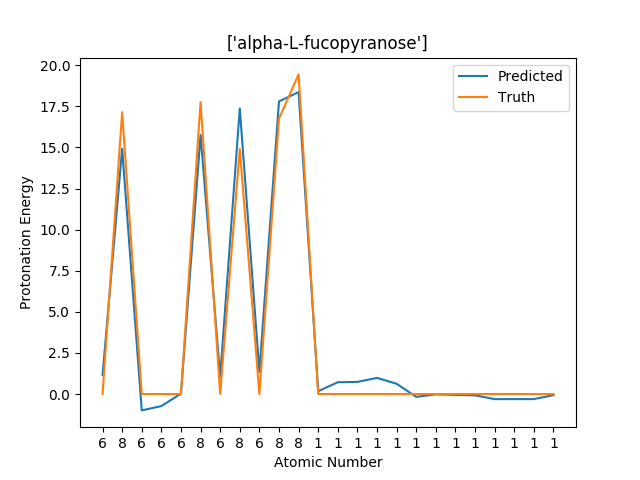}}
\subfigure{\includegraphics[width=0.33\textwidth]{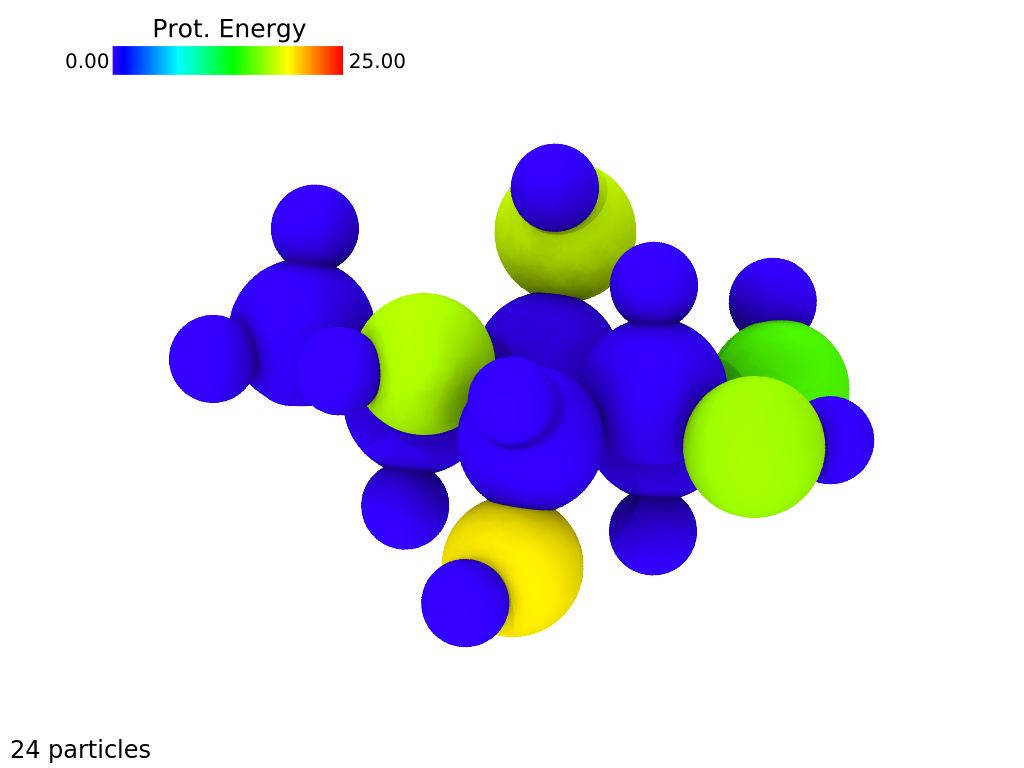}}
\subfigure{\includegraphics[width=0.33\textwidth]{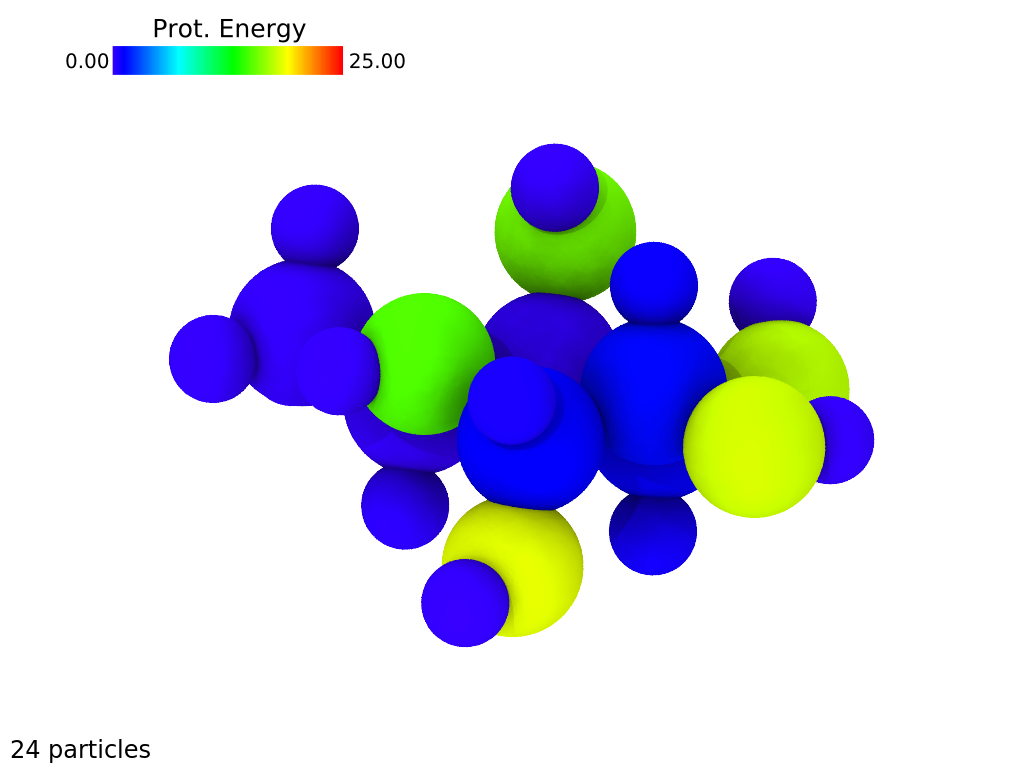}}
} \\

\begin{center}
Alpha-L-fucopyranose
\end{center}

\mbox{
\subfigure{\includegraphics[width=0.33\textwidth]{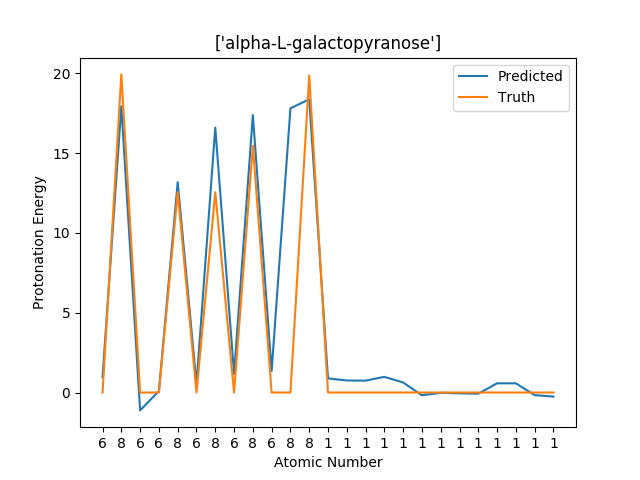}}
\subfigure{\includegraphics[width=0.33\textwidth]{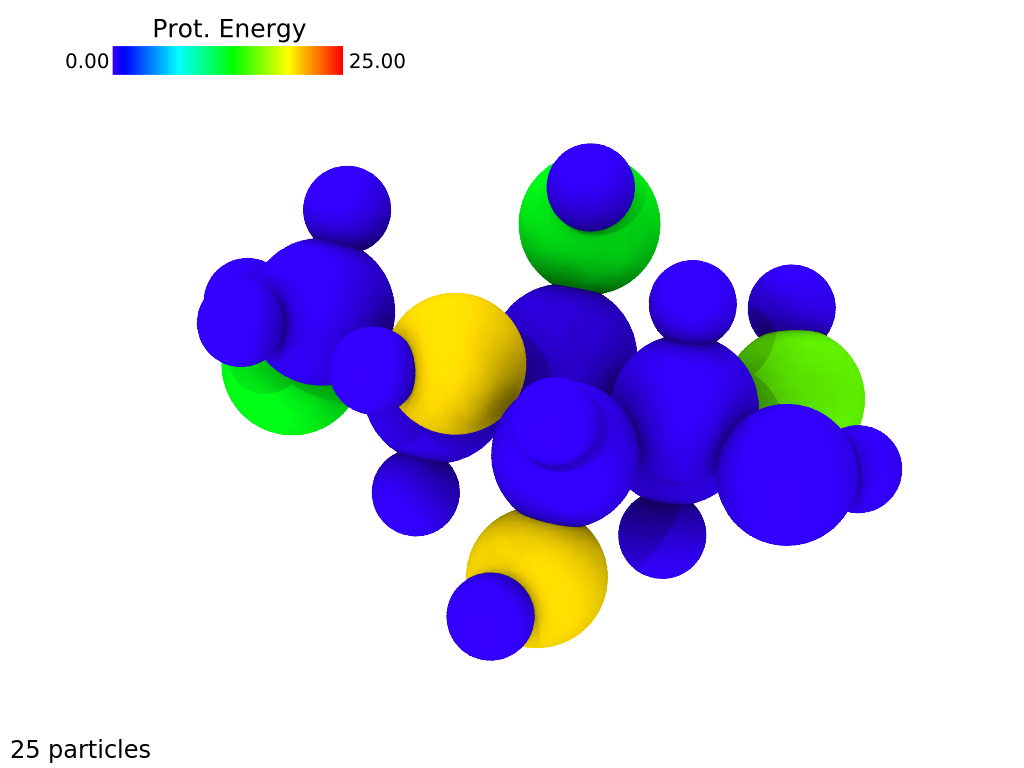}}
\subfigure{\includegraphics[width=0.33\textwidth]{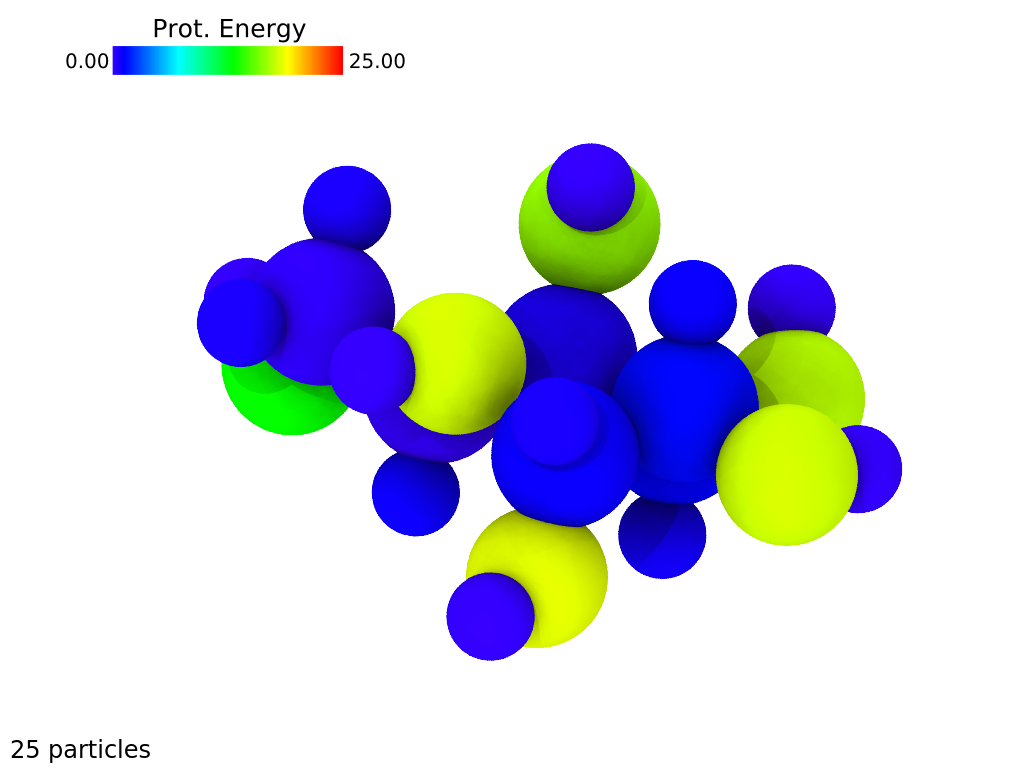}}
} \\

\begin{center}
Alpha-L-galactopyranose
\end{center}

\mbox{
\subfigure{\includegraphics[width=0.33\textwidth]{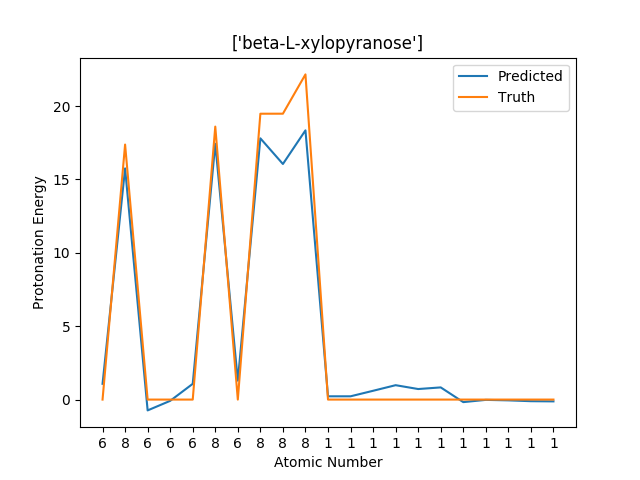}}
\subfigure{\includegraphics[width=0.33\textwidth]{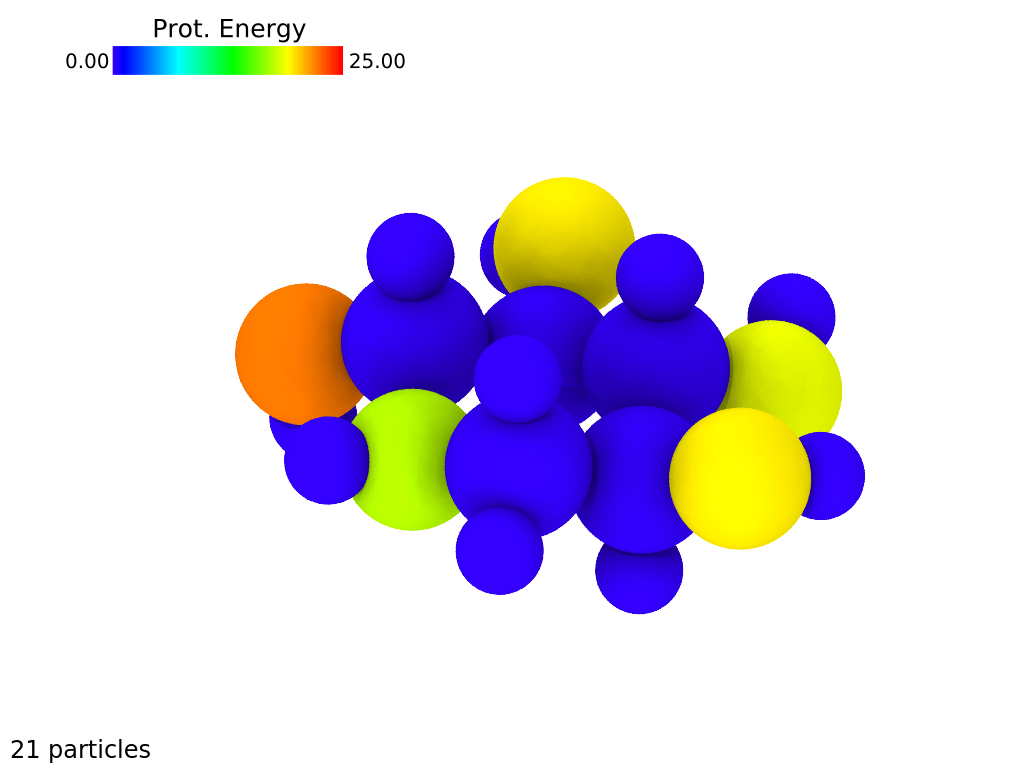}}
\subfigure{\includegraphics[width=0.33\textwidth]{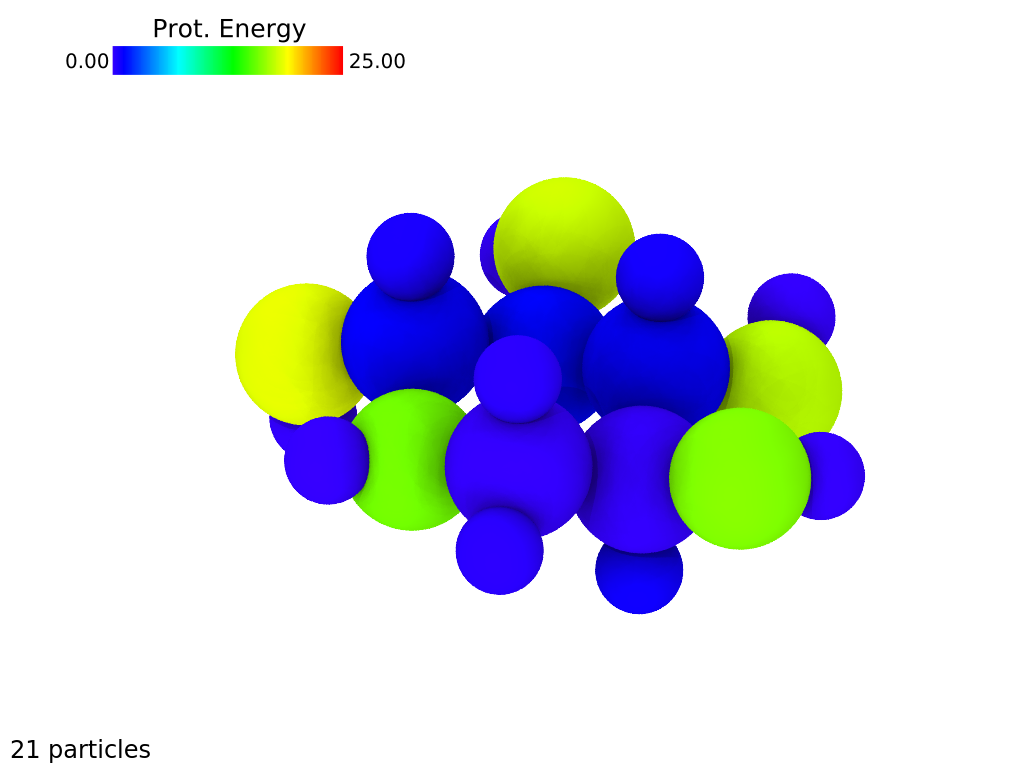}}
} \\

\begin{center}
Beta-L-xylopyranose
\end{center}

\caption{Performance of the proposed method in protonation energy prediction for large organic biofuel molecules. The left column shows predictions of true and predicted protonation energy magnitudes plotted against atomic numbers, the middle column contains graphical representations of the molecules with the true data, and the right column shows results from our method. Note that these molecules are from the testing data set.}
\label{Figure1}
\end{figure}

Figure \ref{Figure1} shows the results for protonation site and energy value prediction for certain large molecules. The plots show magnitudes predicted by the trained GCN superimposed on the truth in addition to 3D visualizations of the molecular structures colored by the value of the protonation energies. We note that these results are for molecules that were a part of the testing dataset (i.e., they are not used for training). In general, we observe that GCN is able to predict trends in energy magnitudes appropriately in addition to finding protonation sites accurately. The sparsity-based RMSE of the testing data was 819.7. Inference speeds of approximately 0.15 seconds per molecule were negligible in comparison with the aforementioned quantum chemical calculations.

\section{Conclusions}

The extreme cost of simulating the characteristics of large organic molecules limits the efficient exploration of trends related to various molecular configurations for various chemistry applications. In this study, we  address this limitation of traditional methods by devising a graph neural network approach to predict properties of a large molecule at the atomic level. 
We developed a graph neural network approach that learns from the high-fidelity G4MP2 bio oil database, and we used the training model to predict the protonation energies of oxygenates, which is crucial to exploring the large space of chemcial transformations required to develop new value-added molecules from naturally abundant chemical species. 
The method outlined here can be utilized for further study of reactivity trends of bio-oil components. Our overarching goal for this work is to study its coupling with molecular discovery concepts in order to quickly assess the viability of a sampled configuration for efficient exploration of the vast space of potential molecules---inaccessible to state-of-the-art methods deployed today.

\subsubsection*{Acknowledgments}
This material is based upon work supported by the U.S. Department of Energy (DOE), Office of Science, Office of Advanced Scientific Computing Research, under Contract DE-AC02-06CH11357. This research was funded in part and used resources of the Argonne Leadership Computing Facility, which is a DOE Office of Science User Facility supported under Contract DE-AC02-06CH11357. This paper describes objective technical results and analysis. Any subjective views or opinions that might be expressed in the paper do not necessarily represent the views of the U.S. DOE or the United States Government.

\clearpage

\section*{References}

[1]  L. A. Curtiss, P. C. Redfern, \& K. Raghavachari, Gaussian-4 theory, \textit{Journal of Chemical Physics}, 126 (2007) 084108. 

[2] L. A. Curtiss, P. C. Redfern, and \& K. Raghavachari, Gaussian-4 theory using reduced order perturbation theory, \textit{Journal of Chemical Physics}, 127 (2007) 124105.

[3] Duvenaud, D. K., Maclaurin, D., Iparraguirre, J., Bombarell, R., Hirzel, T., Aspuru-Guzik, A., \& Adams, R. P. (2015). Convolutional networks on graphs for learning molecular fingerprints. In {\it Advances in neural information processing systems} (pp. 2224--2232).

[4] Xie, T., \& Grossman, J. C. (2018). Crystal graph convolutional neural networks for an accurate and interpretable prediction of material properties. \textit{Physical Review Letters}, \textbf{120(14)}, 145301.

[5] Chen, C., Ye, W., Zuo, Y., Zheng, C., \& Ong, S. P. (2019). Graph networks as a universal machine learning framework for molecules and crystals. \textit{Chemistry of Materials}, \textbf{31(9)}, 3564-3572.

[6] Jin, W., Coley, C., Barzilay, R., \& Jaakkola, T. (2017). Predicting organic reaction outcomes with Weisfeiler-Lehman network. In \textit{Advances in Neural Information Processing Systems}, 2607--2616.

[7] Lei, T., Jin, W., Barzilay, R., \& Jaakkola, T. (2017, August). Deriving neural architectures from sequence and graph kernels. In \textit{Proceedings of the 34th International Conference on Machine Learning}, \textbf{70}, 2024--2033.

\end{document}